\documentclass[a4paper]{jpconf}
\usepackage{graphicx}
\usepackage{float}
\usepackage{wrapfig}

\usepackage{bm}
\usepackage{citesort}
\usepackage{threeparttable}
\usepackage[dvipsnames]{xcolor}

\usepackage[normalem]{ulem}  % \sout{old text} for strikeout

\newcommand{\HI}{H\,{\sc i}}
\newcommand{\HII}{H\,{\sc ii}}
\newcommand{\HeI}{He\,{\sc i}}
\newcommand{\HeII}{He\,{\sc ii}}
\newcommand{\HeIII}{He\,{\sc iii}}
\usepackage[textwidth=1in]{todonotes}

\begin{document}
\title{Estimation of the temperature-density relation in the intergalactic medium at $\mathbf{z\sim 2-4}$ via Ly$\mathbf{\alpha}$ forest}

\author{K N Telikova, S A Balashev and
P S Shternin}

\address{Ioffe Institute, 26 Politeknicheskaya st., St.\ Petersburg, 194021, Russia}

\ead{ks.telikova@mail.ru}

\begin{abstract}

Quasar spectra provide a unique opportunity to investigate the intergalactic medium at high redshifts. The spectral analysis of the Ly$\alpha$ forest allows us to constrain the effective equation of state of the intergalactic medium and its changes during the evolution of the Universe. 
Based on the Voigt profile fitting of Ly$\alpha$ forest lines in 50 high-resolution QSO spectra obtained at Keck telescope
we present new measurements of the power-law index $\gamma$ of  temperature-density relation in IGM for six redshift bins in the range $z=1.6-3.7$. We find that the IGM state is close to isothermal ($\gamma\approx 1$) at $z\sim 3$ which may indicate that HeII reionization occurred at this redshift.

\end{abstract}

\section{Introduction}
One of the widely used methods for probing the intergalactic medium (IGM)  thermal state is based on an analysis of the so-called Ly$\alpha$ forest in the quasar (QSO) spectra. Ly$\alpha$ forest is a composition of absorption features associated with neutral hydrogen (\HI) clouds in the IGM. Observed \HI\ lines are usually decomposed into the individual components, all of which are described by three parameters: the column density of the \HI\ atoms $N$, the Doppler parameter $b$ and the redshift of the absorption system $z$. Doppler parameter $b$ describes the line broadening due to the finite line-of-sight velocity distribution of the atoms in the cloud.
Observed distribution of Ly$\alpha$ forest lines extracted from QSO spectra has a prominent lower envelope in the $(N - b)$ plane \cite{Schaye1999}, which is attributed to a broadening resulting from the pure thermal motions
\begin{equation}\label{eq:minb}
  \mathrm{min}\; b(N)=b_{\rm{th}} \equiv \sqrt{2k_{\rm B}T/m},
\end{equation}
where $k_{\rm B}$ is the Boltzmann constant and $m$ is the hydrogen atom  mass, while
the total broadening of the absorption lines results from a superposition of thermal and peculiar (also called turbulent) motions.
 
Measurements of the  ($N-b$) distribution cutoff allow one to probe the effective equation of state (EOS) of the IGM. In short, the idea is as follows. Under assumptions of uniform background radiation and local hydrostatic equilibrium in the cloud, the column density of the {\HI} absorber that has volumetric density $\rho$ can be determined as \cite{Schaye2001,Rudie2012} 
\begin{equation}\label{eq:rho-N}
	N = 1.3\times 10^{14}\, \Delta^{3/2}\frac{T^{-0.22}}{\Gamma_{-12}}\left(\frac{1+z}{3.4}\right)^{9/2}~{\rm cm}^{-2},
\end{equation}
where $\Delta=\rho/\bar{\rho}$ is the local overdensity, $\bar{\rho}$ is the mean density of the Universe and  $\Gamma_{-12}$ is the hydrogen photoionization rate in units of $10^{-12}$ s$^{-1}$. On the other hand, the relation between the temperature and overdensity, also called the effective EOS, in the low-density IGM after the reionization has a form \cite{HuiGnedin}
\begin{equation}\label{eq:eos}
  T= T_0 \Delta^{\gamma-1},
\end{equation}
where $T_0$ is the temperature at the mean density. 
From eqs~(\ref{eq:minb})--(\ref{eq:eos}) it follows that 
\begin{equation}\label{eq:b-N}
 b_{\rm{th}}= b_0\left(\frac{N}{10^{12}~{\rm cm}^{-2}}\right)^{\Gamma-1} \left(\frac{1+z}{3.4}\right)^{-9(\Gamma-1)/2}, 
\end{equation}
where $b_0$ is the normalization constant, which depends on $T_0$ and $\Gamma_{-12}$, and 

\begin{equation}\label{eq:Gamma-gamma}
    \quad \Gamma-1 = \frac{\gamma-1}{3-0.44(\gamma-1)}.
\end{equation}
 This suggests that the lower envelope of the $(N - b)$ distribution should have the power-law form that relates to the EOS parameters. This technique was frequently applied to different QSO samples resulting in the measurement of EOS at $z\sim2-4$ \cite{Schaye1999,Bolton2014,Hiss2017}.
 In our  previous study \cite{telikova2018} based on the analysis of nine high-resolution QSO spectra we constrained the power-law index $\gamma=1.53\pm 0.07$ (1$\sigma$ confidence) at the mean redshift $z=2.35$. Here we extend this analysis by substantially increasing the number of the QSO spectra which allowed us to follow the evolution of the effective EOS parameters with redshift.

%%%%%%%%%%%%%%%%%%%%%%%%%%%%%%%%%%%%%%%%%%%%%%%%%%%%%%%%%%%%%%%%%%%%%%%%%%%%%%%%%%%%%%%
\section{Analysis}
\begin{figure}[th]
\centering
\includegraphics [width=0.9\textwidth]{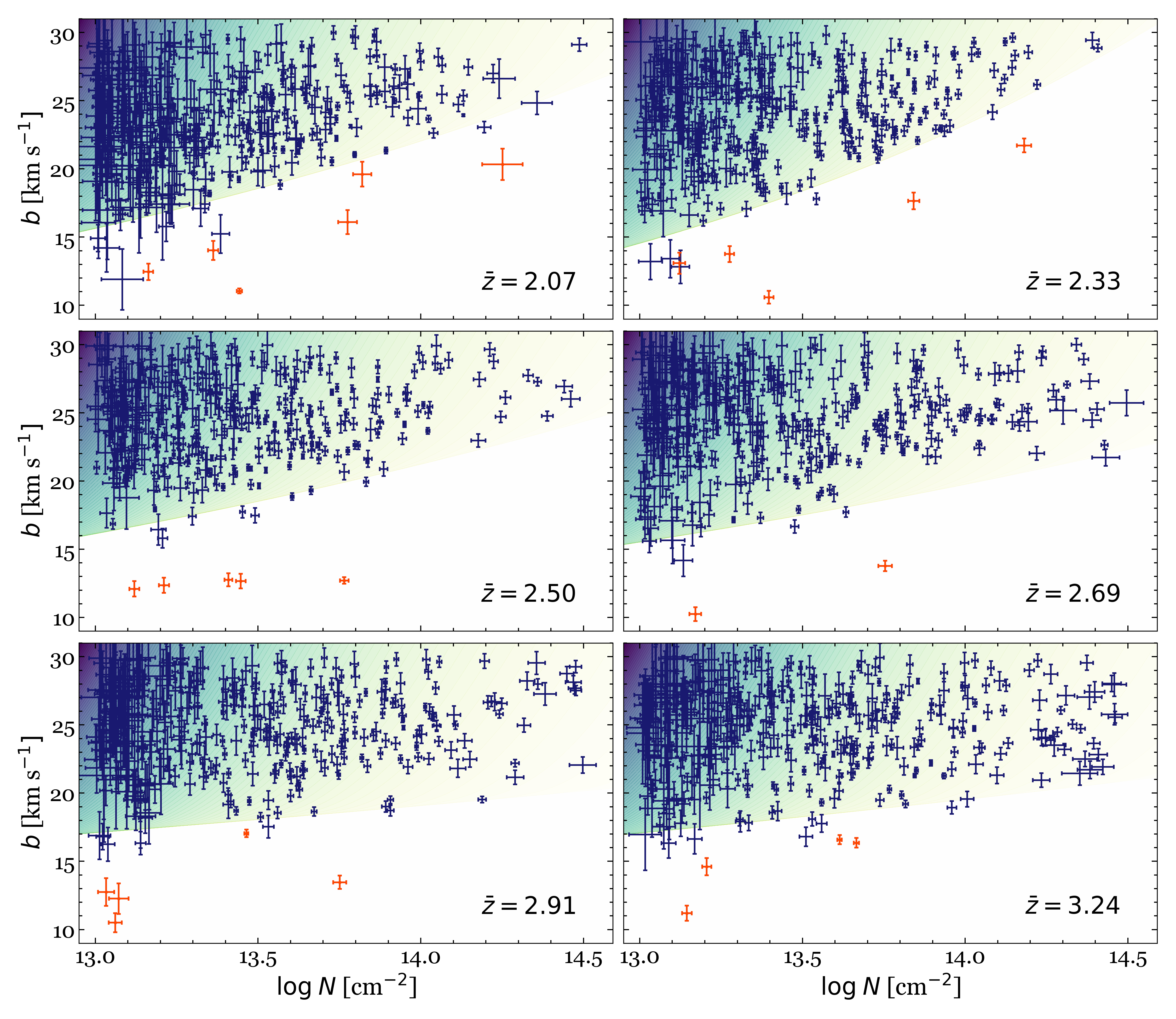}
\caption{Crosses: samples of the Ly$\alpha$ forest lines in six redshift bins, after the metal rejection. Mean bin redshifts are shown in each panel. Gradient-filled regions show best fits of the probability density distribution functions. Points coloured in red most likely are outliers, see text for details.}
\label{fig:data_fit}
\end{figure}

%%%%%%%%%%%%%%%%%%%%%%%%%%%%%%%%%%%%%%%%%%%%%%%%%%%%%%%%%%%%%%%%%%%%%%%%%%%%%%%%%%%%%%%
%%%%%%%%%%%%%%%%%%%%%%%%%%%%%%%%%%%%%%%%%%%%%%%%%%%%%%%%%%%%%%%%%%%%%%%%%%%%%%%%%%%%%%%
\begin{center}
\begin{table}[t]
\centering
\caption{\label{tab:pars} Fit parameters for six redshift bins. }\lineup
\begin{tabular}{@{}l*{15}{l}}
\br
\rule{0pt}{1ex}  Redshift range&$1.62 - 2.22$&$2.22-2.42$&$2.42-2.58$&$2.58-2.80$&$2.80-3.03$&$3.03-3.74$\\
\rule{0pt}{1ex}  Mean redshift&$2.07$&2.33&2.50&2.69&2.91&3.24\\
\rule{0pt}{1ex}  
$\Gamma-1$&$0.15^{+0.02}_{-0.01}$&$0.20^{+0.01}_{-0.02}$&$0.12^{+0.02}_{-0.02}$&$0.10^{+0.02}_{-0.02}$&$0.05^{+0.02}_{-0.05}$&$0.06^{+0.02}_{-0.01}$\\
\rule{0pt}{2.6ex}
$\gamma-1$&$0.42^{+0.05}_{-0.04}$&$0.56^{+0.04}_{-0.05}$&$0.34^{+0.07}_{-0.07}$&$0.30^{+0.06}_{-0.05}$&$0.14^{+0.05}_{-0.14}$&$0.17^{+0.05}_{-0.04}$\\
\rule{0pt}{2.6ex}  
$\log b_0$&$1.01^{+0.02}_{-0.04}$&$0.95^{+0.03}_{-0.03}$&$1.10^{+0.03}_{-0.04}$&$1.10^{+0.03}_{-0.03}$&$1.20^{+0.09}_{-0.03}$&$1.20^{+0.01}_{-0.02}$\\
\mr
\rule{0pt}{2.6ex}
$p$&\multicolumn{6}{c}{\m$1.18^{+0.08}_{-0.09}$} \\
\rule{0pt}{2.6ex}  $\beta$&\multicolumn{6}{c}{$-1.73^{+0.03}_{-0.04}$}\\
\br
\end{tabular}
\end{table}
\end{center}
%%%%%%%%%%%%%%%%%%%%%%%%%%%%%%%%%%%%%%%%%%%%%%%%%%%%%%%%%%%%%%%%%%%%%%%%%%%%%%%%%%%%%%%
We used 50 QSO spectra with high resolution ($\sim 36000-72000$) and signal-to-noise ratio ($\sim 20-100$) from the KODIAQ (Keck Observatory Database of Ionized Absorption toward Quasars) survey \cite{OMeara2017}. We updated our automatic routine \cite{telikova2018} and based the search on the cross-correlation between an observed spectrum and model Ly$\alpha$ Voigt profiles. Synthetic Ly$\alpha$ lines were calculated on a dense $(N, \,b)$ grid. Grid steps for each spectrum were estimated from the Fisher matrix calculations accounting for the resolution and signal-to-noise ratio. 
For each grid point, cross-correlation was calculated along the redshift axis in a range between Ly$\alpha$ and Ly$\beta$ QSO emission lines.  Peaks in the cross-correlation function indicate the positions of the Ly$\alpha$ forest lines, and the Fisher-matrix-based grid resolution ensures that lines are not missed. As a final step, the line parameters were refined by the least squares fit and the uncertainties $\sigma^N$ and $\sigma^b$ were obtained from the $\chi^2$ likelihood confidence intervals. Such procedure allows to select and fit only solitary Ly$\alpha$ lines. Afterwards, we mitigated the selection criteria to allow one of the line wings to be partially blended, that led to the increase in the sample volume by a factor of two. 
The final sample of Ly$\alpha$ lines was cleaned from the metal lines by visual inspection. The metal absorption lines in quasar spectra usually correspond to the doublet lines or can be associated with damped Ly$\alpha$ systems.  
Therefore we masked all the lines for which we found evident counterparts with similar velocity structure. 
After cleaning, the sample contained 2268 individual absorption systems in the redshift range $(1.6-3.7)$. We divided it into 6 redshift bins (first two rows in table~\ref{tab:pars}) containing approximately the same number of lines. 
The corresponding samples are shown in figure~\ref{fig:data_fit} with blue error crosses. 

We fit the obtained absorption systems distribution $(N, \,b)$ by a model distribution function. We assume the following distribution form 
\begin{equation}\label{eq:pdf_N_b}
  f(N,\, b)=\int f_N(N)f_{\rm turb}(b_{\rm{turb}})\delta \left(b-\sqrt{b_{\rm{th}}^2+b_{\rm{turb}}^2} \right)\rm{d} \emph{b}_{\rm{turb}},
\end{equation}
where $b_\mathrm{turb}$ is the broadening due to turbulent motions, $f_N(N)$ and  $f_{\rm turb}(b_{\rm{turb}})$  are column density distribution and turbulent broadening parameter distribution, respectively. Delta function in eq~(\ref{eq:pdf_N_b}) specifies that the thermal and turbulent motions are uncorrelated (microturbulence assumption) and corresponding broadening parameters add in quadrature to give the total $b$.
It is known from observations that to a good approximation column density distribution obeys the power-law shape $f_N(N)\propto N^\beta$ in the $N$ range considered here \cite{Janknecht2006J,Rudie2013}. In order to select the shape of the $f_\mathrm{turb}(b_\mathrm{turb})$ distribution, we first visually estimated the putative cutoff position $b_\mathrm{th}(N)$ in the obtained sample. 
 After that we plotted the distribution of the absorption systems over $b_{\rm turb}$ and noted that within grid limits ($b = 10-30$ km~s$^{-1}$) it also has approximate power-law shape. Therefore we choose the distribution over the turbulent broadening parameter in the form $f_{\rm turb}(b_{\rm{turb}})\propto b_{\rm{turb}}^p$. 
The advantage of this approach is that more information contained in the sample is used 
in comparison to previous studies, where the low boundary position was determined based on various iterative rejection algorithms or  where some integral statistics of the obtained sample was used  \cite{Schaye1999,Ricotti2000,Bolton2014,Hiss2017}.

Visual inspection of the metal-cleaned samples showed that some lines have $b$ values much smaller than the estimated thermal one, that can not be attributed to the measurement uncertainties. Some of them can correspond to unidentified metal lines, for instance when their putative counterparts are completely blended with saturated Ly$\alpha$ forest lines. Additionally, nonuniform background radiation and special conditions can result in peculiar effective EOS for a particular cloud \cite{Puchwein2018}. Therefore, these lines can not be described in the framework of the adopted model and are considered outliers.

The parameters of the model were constrained under the Bayesian scheme that takes into account the presence of the outliers. Specifically, the likelihood function for the $i$'th observed data point to be generated from our model is
\begin{equation}\label{eq:L_data}
{\cal L_{\rm data}}(N_i,\, b_i) =  \frac{\int f(\widetilde{N}, \widetilde{b}) \exp(-\frac{(\widetilde{N}-N_i)^2}{2\sigma_{N_i}^2}) \exp(-\frac{(\widetilde{b}-b_i)^2}{2\sigma_{b_i}^2}) \,{\rm d}\widetilde{N}\,{\rm d}\ \widetilde{b}}
{\int f(\widetilde{N}, \widetilde{b})I(N, b) \exp(-\frac{(\widetilde{N}-N)^2}{2\sigma_{N_i}^2}) \exp(-\frac{(\widetilde{b}-b)^2}{2\sigma_{b_i}^2}) \,{\rm d}N\,{\rm d}b\,{\rm d}\widetilde{N}\,{\rm d}\widetilde{b}},
\end{equation}
where the generative model (\ref{eq:pdf_N_b}) is convoluted with Gaussian functions to account for the measurement errors. The normalization in the denominator in eq~(\ref{eq:L_data}) takes into account that our sample is truncated. Here $I(N,b)$ is the indicator function that is equal to one if an $(N,\, b)$ pair falls in the specified box and is zero otherwise. The outliers are included following the mixture model receipt \cite{Hogg2010} that assumes that each data point has a probability $P_b$ to be generated from the unknown bad points distribution with likelihood ${\cal L_{\rm out}}(N_i,\, b_i)$ instead of being generated from the model distribution. Following \cite{Hogg2010} we took the normal distribution for outliers with some mean $Y_b$ and variance $V_b$, although the other choices are possible. Notice that the outlier likelihood ${\cal L_{\rm out}}(N_i,\, b_i)$ accounts for the measurement errors and sample truncation in the same way as ${\cal L}_\mathrm{data}$. The final likelihood function under the mixture model is
\begin{equation}\label{eq:L}
    {\cal L} =\prod\limits_{i} \left[ (1-P_b) {\cal L_{\rm data}}(N_i,\, b_i) + P_b{\cal L_{\rm out}}(N_i,\, b_i) \right].
\end{equation}

We fit the model simultaneously in 6 redshift bins. Two parameters ($\Gamma$,\, $\log b_0$) were fitted for individual data bins, while $\beta$, $p$, $P_b$ and parameters of outlier distribution were estimated for the full $(N,\, b)$ sample. This results in 17 fit parameters in total. Flat priors were used on all parameters except $V_b$ where the flat prior on its logarithm was used. The posterior was sampled using the affine Markov Chain Monte Carlo  (MCMC) sampler \textsf{emcee} \cite{Foreman-Mackey2013}. 
Fit results for the parameters of interests are summarized in table~\ref{tab:pars}. Uncertainties  are calculated using credible intervals of the marginalized posterior distributions at 0.683 confidence level. Best-fit $(N-b)$ distributions are shown by the gradient-filled areas in figure~\ref{fig:data_fit}. Red points mark the most probable outliers, i.e. the points for which the posterior-based expectation value of the second term in eq~(\ref{eq:L}) 
is larger than those of the first term. We see that the number of outliers is small, however, being not accounted for they can significantly hamper the cutoff determination. 

%%%%%%%%%%%%%%%%%%%%%%%%%%%%%%%%%%%%%%%%%%%%%%%%%%%%%%%%%%%%%%
\begin{figure}[t]
\begin{center}
\includegraphics [width=0.8\textwidth]{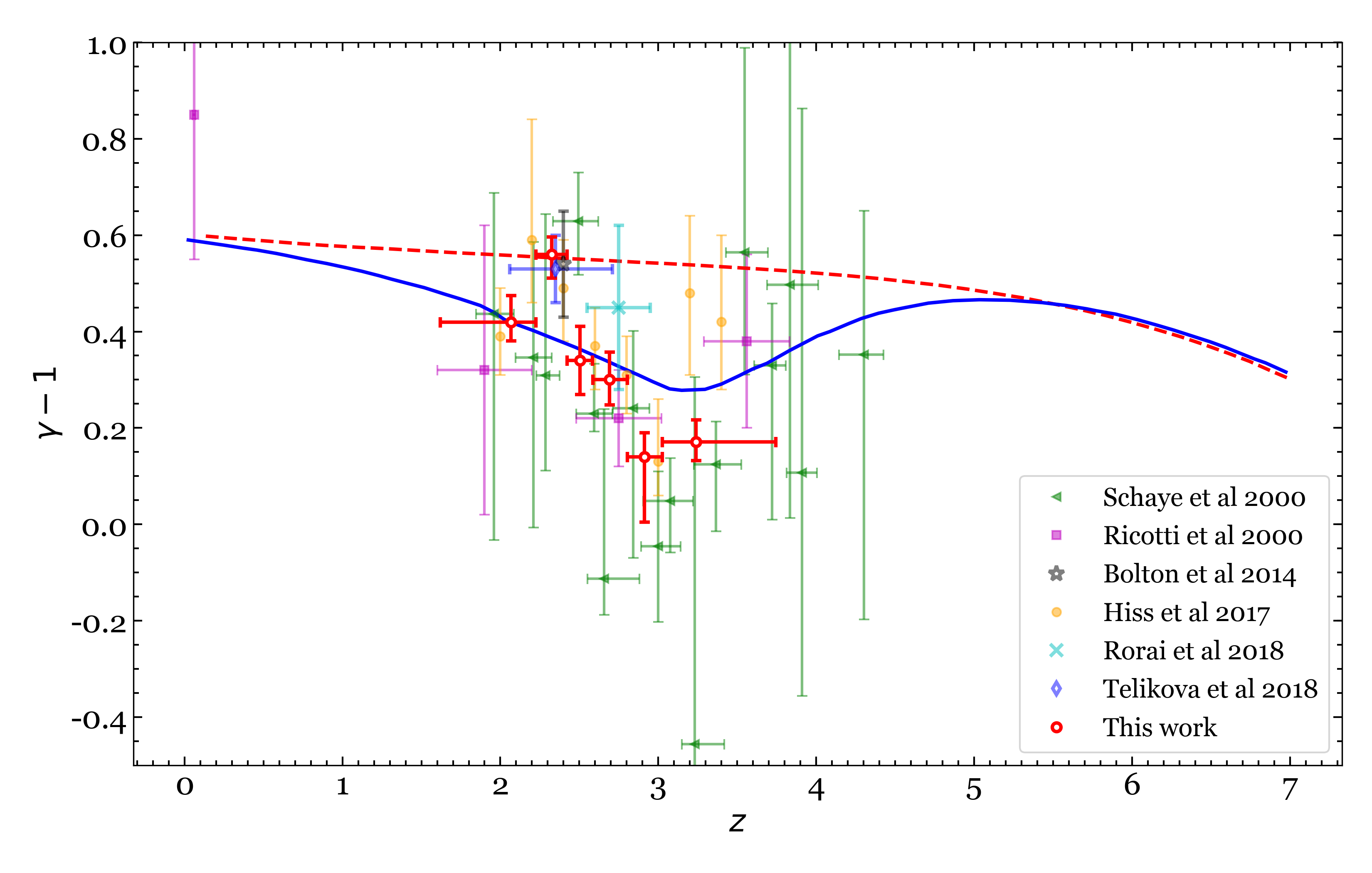}
\caption{Evolution of $\gamma$ with $z$. Error crosses show measurements via Ly$\alpha$ forest \cite{Schaye2000,Ricotti2000,Bolton2014,Hiss2017,Rorai2018,telikova2018}. Results of the present work are shown with open circles.
Blue solid and red dashed curves are models of thermal evolution of the IGM from \cite{UptonSanderbeck2016} with and without \HeII\ reionization, respectively.} 
\label{fig:gamma}
\end{center}
\end{figure}
%%%%%%%%%%%%%%%%%%%%%%%%%%%%%%%%%%%%%%%%%%%%%%%%%%%%%%%%%%%%%%%

\section{Discussion and conclusions}
Based on the inferred values of $\Gamma-1$ we find the temperature-density relation index $\gamma-1$ from eq~(\ref{eq:Gamma-gamma}). The results in six bins are plotted in figure~\ref{fig:gamma} with open circles along with various results published in the literature (other symbols). With the  blue open diamond we show the result of our previous work \cite{telikova2018} at $z\approx 2.35$ which is consistent with the present result for the second redshift bin. 
We find the significant drop of $\gamma$ at $z\approx 3$ indicating that the effective EOS at this redshift is close to the isothermal one ($\gamma\approx 1$). This finding is in line with the recent results by Hiss {\it et al} \cite{Hiss2017} (filled orange dots in figure~\ref{fig:gamma}) and older results by Schaye {\it et al} \cite{Schaye2000} (green triangles in figure~\ref{fig:gamma}) although the latter have rather large uncertainties.  

The reason for the drop in $\gamma$ can be understood as follows. It is currently believed that the IGM has encountered two major reioniziation events at the intermediate redshifts driven by the background emission from the first galaxies and QSOs. The first event is the hydrogen reionization \HI\ $\to$ \HII\  which completed at $z\sim6$. Approximately at the same time, the first helium reionzation \HeI\ $\to$ \HeII\ is thought to had occurred. Long after the \HI\ reionization the balance between heating and cooling processes set the effective EOS in the form (\ref{eq:eos}). If the main heating source is the residual neutral hydrogen photoheating then $\gamma\approx 1.6$ is predicted \cite{HuiGnedin}.
The presence of the dip in $\gamma(z)$ at $z\approx 3$ can be attributed to a second helium reionozation event \HeII\ $\to$ \HeIII, which was driven by QSOs. This process led to additional heating of the gas that was mostly independent of the overdensity resulting therefore in isothermalization of the IGM. According to theoretical simulations, this event indeed took place at $z\sim 2-3$.
This is illustrated with the dashed and solid lines in figure~\ref{fig:gamma} which show the  simulation results from \cite{UptonSanderbeck2016} with and without \HeII\ reionization, respectively. 

Nevertheless, typical uncertainties for the most of the measurements in figure~\ref{fig:gamma} are large and do not allow to prove the significance of the dip. For instance, Hiss {\it et al}~\cite{Hiss2017} point out that their results are in fact consistent with the constant $\gamma=1.4$ throughout the whole studied redshift range. Our data have smaller formal statistical errors (table~\ref{tab:pars}) and do not allow for a flat $\gamma(z)$ dependence. Nevertheless, in agreement with \cite{Hiss2017}, we do not find an inverted ($\gamma<1$) EOS even at the bottom of the dip.

In principle, the position of the cutoff in the ($b-N$) distribution allows one to constrain not only the $\gamma$ parameter, but also the temperature at the mean density $T_0$ [see eq~(\ref{eq:eos})] based on the inferred value of the intercept parameter $b_0$. This is not a straightforward task since, according to eq~(\ref{eq:rho-N}), EOS depends on the unknown hydrogen ionization rate $\Gamma_{-12}$.
This can be dealt with if independent measurements of $\Gamma_{-12}$ are available. For instance, measurements of the mean opacity of the Ly$\alpha$ forest constrain the ratio $T_0^{-0.7}/\Gamma_{-12}$ \cite{Faucher2008}. Together with $b_0$ measurements this makes possible to estimate both $T_0$ and $\Gamma_{-12}$. However, the results of \cite{Faucher2008} were criticized in refs \cite{Bolton2014} and \cite{Hiss2017} as they give too small $\Gamma_{-12}$. Hiss {\it et al} \cite{Hiss2017} suggested to use the values of $N$ at the mean density ($\Delta=1$) based on the observation that $N(\Delta=1)$ has little scatter in simulations carried over a large grid of parameters provided $\Gamma_{-12}$ is adjusted to give the correct mean Ly$\alpha$ opacity. Two approaches lead to significant differences (by a factor of 1.5) in the inferred values of $T_0$. 

Moreover, simulations reported in \cite{Garzilli2015} show that the cutoff position depends strongly on the inclusion of the pressure smoothing in the model. In contrast,  such a dependence was not found in \cite{Hiss2017}.
This can be attributed to  different methods used to estimate the position of the cutoff in these references, but the exact reasons are unclear. Taking to account these problems we do not discuss $T_0$ inference in this short note and defer its discussion for a future work.

We wish to note that our results may be subject to systematic uncertainties since they rely on the specific parametric forms assumed for the $b_\mathrm{turb}$ distribution and the outlier distribution. Nevertheless we find aposteriori that the power-law distribution over $b_\mathrm{turb}$ agrees with the data fairly well. In the future studies we plan to increase the statistics by increasing the number of QSO spectra and test the robustness of our results to change of the distributions' shapes.

\ack The work was supported by the Russian Science Foundation, grant 18-12-00301. 

%\newpage
\section*{References}
\bibliographystyle{iopart-num}
\bibliography{references.bib}
\end{document}